 \documentclass[final,1p,times]{elsarticle}
\usepackage{amssymb}
\usepackage{longtable}
\usepackage{float}
\usepackage{url}
\usepackage{hyperref}
\usepackage{ragged2e}
\usepackage{booktabs}
\usepackage{makecell}
\usepackage{tabularx}
\usepackage{multirow}

\newcommand{\dis}[0]{{{\sc Discover Stage}}}
\newcommand{\defi}[0]{{{\sc Define Stage}}}
\newcommand{\dev}[0]{{{\sc Develop Stage}}}
\newcommand{\del}[0]{{{\sc Deliver Stage}}}

\journal{International Journal of Human-Computer Studies}

\begin{document}
\begin{frontmatter}

\title{Designing Child-Centric AI Learning Environments: Insights from LLM-Enhanced Creative Project-Based Learning}\corref{cor2}

\author[inst1]{Siyu Zha}

\affiliation[inst1]{organization={Tsinghua University},
            city={Beijing},
           country={China}
            }

\author[inst1]{Yuehan Qiao}
\author[inst1]{Qingyu Hu}
\author[inst2]{Zhongsheng Li}
\author[inst1]{Jiangtao Gong*}\corref{cor1}
\author[inst1]{Yingqing Xu}

\affiliation[inst2]{organization={School of Information, University of Texas at Austin},
            city={Austin},
           country={USA}
            }
\cortext[corl]{Corresponding author}
\cortext[cor2]{Parts of this paper have been accepted by the CHI 2024 Workshop on Child-centred AI Design}

\begin{abstract}
Project-based learning (PBL) is an instructional method that is very helpful in nurturing students' creativity, but it requires significant time and energy from both students and teachers. Large language models (LLMs) have been proven to assist in creative tasks, yet much controversy exists regarding their role in fostering creativity. This paper explores the potential of LLMs in PBL settings, with a special focus on fostering creativity. We began with an exploratory study involving 12 middle school students and identified five design considerations for LLM applications in PBL. Building on this, we developed an LLM-empowered, 48-hour PBL program and conducted an instructional experiment with 31 middle school students. Our results indicated that LLMs can enhance every stage of PBL. Additionally, we also discovered ambivalent perspectives among students and mentors toward LLM usage. Furthermore, we explored the challenge and design implications of integrating LLMs into PBL and reflected on the program. By bridging AI advancements into educational practice, our work aims to inspire further discourse and investigation into harnessing AI's potential in child-centric educational settings.
\end{abstract}


\end{frontmatter}

\section{Introduction}
Project-based learning (PBL) is a well-known instructional model that encourages students to develop creativity by solving complex, real-world problems in groups~\cite{bell2010project}. Some empirical research shows that PBL significantly arouses learning interest, stimulates students‘ creativity, and promotes active student learning~\cite{kokotsaki2016project,hernandez2009learning,genc2015project}.
However, PBL also presents numerous challenges, as indicated by various empirical studies, students often struggle with managing project schedules, particularly when it comes to balancing in-depth exploration and creation with timely delivery~\cite{hung2012project, doppelt2003implementation}. Moreover, teachers must consciously navigate and manage complex relationships between students, teachers, and tasks in PBL\cite{wurdinger2007qualitative, aksela2019project}. They need to pay more attention to providing suitable guidance to ensure educational coherence and quality~\cite{zhou2012problem, english2013supporting}. With the rapid development of artificial intelligence technology, large language models (LLMs) have contributed to many powerful tools, such as ChatGPT\footnote{\url{https://chat.openai.com}}, New Bing\footnote{\url{https://www.bing.com}}, and Bard\footnote{\url{https://www.bard.google.com}}. The rise of LLMs has sparked an extensive discourse among educators and child psychologists, particularly in its application in child-centered AI design. Emerging digital technologies make it easier for students to engage in the process of designing and developing their projects, as they can document the entire process and share their creations in a digital format rapidly~\cite{patton2012work,d2021intensive,braga2022research}. Several studies have begun exploring their support for creativity due to their human-like intelligence, reasoning, and natural interactivity~\cite{swanson2021story,organisciak2023beyond,huang2023causalmapper}. For instance, LLMs can support creative writing~\cite{swanson2021story}, assist professional designers in constructing and exploring design space~~\cite{huang2023causalmapper}, score the divergent thinking automatically~\cite{organisciak2023beyond}, among other applications.

Despite LLM-powered tools having the potential to profoundly impact creativity and other educational goals~\cite{,tan2023towards,opara2023chatgpt}, 
Tan, K. et al. investigated strategies for applying LLMs to classroom teaching, exploring the potential of AI to enhance teacher-student dialogues and improve teaching quality~\cite{tan2023towards}. Urban, M. et al. found that ChatGPT can improve university students' creative problem-solving performance~\cite{urban2023can}. Other researchers focused on their use for idea generation in innovation~\cite{girotra2023ideas,ekvall2023integrating,bilgram2023accelerating}. Amidst these studies, special attention is given to LLMs' practical application in education, particularly in PBL scenarios~\cite{wan2023felt,salas2022artificial}. Advocates believe mastering LLM-powered tools can enhance students' creativity and problem-solving skills~\cite{bitzenbauer2023chatgpt,hadi2023large}. Moreover, it's found that PBL, which emphasizes creative solutions to real-world problems, significantly promotes students' overall development~\cite{zhai2022chatgpt}. Therefore, the exploration of integrating LLMs into creative PBL is particularly important. Based on this, we have proposed the following research questions:

\begin{itemize}
 \item How can LLMs support the needs of various stages in PBL?
 \item What are the attitudes of students and mentors towards the integration of LLMs into PBL in actual projects?
 \item What challenges do LLMs encounter in supporting the PBL process?
 \item What design considerations arise from the use of LLMs in supporting the PBL process?

\end{itemize}

In this study, we first conducted an exploratory study engaging 12 students to understand the use of LLM in creative PBL and identified five design considerations. Based on our exploratory results, we then designed an instructional experiment involving 31 middle school students that illustrated how LLMs can boost every stage of creative PBL through qualitative analysis. We distilled insights around students’ complex and, sometimes, conflicted attitudes toward the effective use of LLMs. We then further reflected on the design considerations and challenges of implementing LLM-powered creative PBL. The contributions of this study are as follows:
i) Successfully implemented a LLM-powered PBL, observing and summarizing the support provided by LLMs at each stage of PBL;
ii) Conducted in-depth interviews with students and teachers to explore their complex attitudes towards the integration of LLMs into educational settings;
iii) Identified and summarized the challenges associated with incorporating LLMs into the PBL process;
iv) Concluded with design considerations and implications for incorporating LLMs into the PBL process.
Overall, this study provides a valuable practical exploration of the application of LLMs and insights for future research regarding their use in relevant applications.

\section{Related Work}

\subsection{Project-Based Learning in the Classroom}
PBL is an innovative approach that involves students in active inquiry and collaboration, which sharpens their creative skills and problem-solving skills through real-world scenarios~\cite{bell2010project, jeon2023developing}. It also motivates interdisciplinary learning by incorporating multiple academic subjects in the classroom~\cite{massoud2023interdisciplinary}. In summary, PBL is a student-centered instructional method characterized by student autonomy, constructive investigations, goal-setting, collaboration, communication, and reflection within real-world practices. It's implemented in various contexts and across different levels of education, from primary to higher education~\cite{kokotsaki2016project}.

Previous research has established that PBL is a widely recognized and practiced teaching method in classrooms~\cite{hernandez2009learning}.
Genc M. et al. revealed that using PBL in the classroom is beneficial, which can enhance creativity, encourage research, and provide permanent learning~\cite{genc2015project}. In a case study, Novak et al. explored a semester-long 7th-grade PBL curriculum on water quality. They discovered that PBL engages students in investigating meaningful questions relevant to their everyday lives~\cite{novak2019case}. Similarly, Asan et al. presented a design of an effective computer PBL class. They found that PBL provides the opportunity for students to apply theoretical and practical knowledge while developing their group work and collaboration skills in the classroom~\cite{asan2005implementing}. 

However, implementing PBL in classrooms presents several challenges. Students need to develop collaboration, innovation, and critical reasoning abilities as they jointly investigate and solve problems, learning necessary knowledge and skills along the way~\cite{zhou2012problem,hung2012project,doppelt2003implementation}. It's also challenging to balance in-depth exploration with quick delivery under time limitations and maintaining educational quality and coherence~\cite{patton2012work}. Moreover, PBL requires teachers to support students in group work, planning, problem-solving, idea testing, and peer presentation~\cite{zhou2012problem,english2013supporting}. Guiding students to complete a creative PBL project within a set timeframe also presents a significant challenge for teachers~\cite{wurdinger2007qualitative,aksela2019project}. Therefore, this study aims to explore the potential of engaging new technologies, such as LLMs, to support creative PBL in middle school classrooms.

\subsection{Creative Support Tools in Learning}
Creativity Support Tools (CSTs) play a fundamental role in the study of creativity in Human-Computer Interaction (HCI)~\cite{10.1145/3290605.3300619,gong2021holoboard,gong2014paperlego}. They operate on digital systems to encourage creativity and benefit users at various creative stages~\cite{shneiderman2007creativity,cherry2014quantifying}. CSTs inspire creativity in diverse groups and enhance learning. For instance, YOLO, a robot for children to boost new ideas and promote their creative learning~\cite{10.1145/3078072.3084304}, and ShadowStory, a digital storytelling system fostering creativity and cultural connection among children~\cite{10.1145/1978942.1979221}. In addition, the concept of an online creative community has also been proposed. This innovative platform encourages learning through the sharing of creative work, thereby fostering a sense of community and collaboration among its members~\cite{10.1145/2998181.2998195}. These examples underscore the vast potential and importance of CSTs in nurturing creativity and enhancing learning experiences in the digital age.

AI's potential in enhancing creativity through random stimuli is increasingly recognized in CSTs ~\cite{beaney2005imagination,figoli2022artificial}. Some studies highlight the effectiveness of 'Generators' category tools, which contain challenging biases and introduce unexpected elements to enrich the creative process. ~\cite{hwang2022too,baidoo2023education}. Davis et al. further discussed 'Generative fashion', a tool promoting divergent and convergent thinking, illustrating the need for tools aligned with design space exploration to fully utilize AI's creative potential, especially in professional creative fields~\cite{davis2023fashioning}. Cautela et al. suggested that the potential application of AI in creative practices is vast but largely untapped, indicating new opportunities in creative AI applications ~\cite{xue2019researcher}. Thus, further exploring and utilizing AI's creative potential in various fields is crucial, especially in students' learning.

In conclusion, CSTs have demonstrated the potential to improve students’ creativity in learning. However, their actual use, especially involving AI in PBL classrooms, remains relatively rare. Our study aims to explore the opportunities and challenges of using LLMs as CSTs in PBL classrooms, with a focus on middle school students. Our goal is to offer insights into how LLMs can be effectively incorporated into educational settings to enhance creative learning.

\subsection{Large Language Models in Education}
The adoption of LLMs in education is progressing cautiously, with methodologies for their effective use still emerging. Recent studies highlight the need for more structured strategies and guidelines for integrating LLMs into educational frameworks ~\cite{salas2022artificial,leaton2020artificial,rudolph2023chatgpt}. This reflects a broader trend where the deployment of LLMs in education remains a work in progress. 
Researchers' views on incorporating LLMs into educational settings vary widely. Some advocate for their potential to transform learning environments, citing tools like ChatGPT as instrumental in enriching education, supporting educators, and redefining objectives and practices \cite{lo2023impact,abd2023large,perkins2023academic,alqahtani2023emergent}. 
For instance, researchers highlighted ChatGPT's role in boosting creative problem-solving and self-efficacy among university students through collaborative creation \cite{tan2023towards,aydin2023chatgpt}. On the other hand, there are growing concerns regarding the potential misuse of LLMs in educational settings. Susnjak et al. revealed that essays generated by ChatGPT display advanced critical thinking and textual discourse, sparking debates about the risks of cheating in online assessments ~\cite{susnjak2022chatgpt}.
These findings all underscore the ethical implications and the need for careful oversight when using LLMs. Despite differing opinions, LLMs are still recognized as a vital support in educational instruction, indicating a significant role for these technologies in future learning ecosystems\cite{vazquez2023chatgpt,nguyen2023ethical}.

LLMs can effectively assist in curriculum exercises to enhance creative learning. Recent studies are shining a light on LLMs, particularly GPT-4, showing that they can be powerful tools to support creativity~\cite{hadi2023survey,chakrabarty2023creativity}. Noy et al. found that ChatGPT can enhance the quality of solutions~\cite{noy2023experimental}. Douglas et al.'s research points out that LLMs can do even better than traditional brainstorming methods and average human performance in specific tasks when given the right prompts and paradigms ~\cite{summers2023brainstorm,yao2023tree}. It's also worth noting that GPT-4 isn’t just great for individual creative thinking. It has a lot to offer when it comes to tackling collaborative challenges ~\cite{liu2023summary,latif2023artificial}. The reason why LLMs can empower students' learning is that they can provide personalized information, resources, and step-by-step solutions based on students’ unique learning needs and interests~\cite{kasneci2023chatgpt}. Therefore, researchers recommended designing AI-involved learning tasks that engage students in real-world problem-solving to meet learning objectives effectively~\cite{woithe2023understanding}. Our study seeks to fill this gap by providing an in-depth analysis of how LLMs, especially in a PBL context, can aid students in solving real-world problems and developing their creativity.

\section{Design LLM-empowered PBL Program}
The rise of LLMs has sparked an extensive discourse among educators. Tools like ChatGPT can assist students in creative problem-solving within a PBL setting. However, selecting suitable LLM tools, designing teaching segments, boosting students' acceptance of new technologies, and requiring additional training for teachers all pose challenges. Therefore, we designed an exploratory study to identify design implementation strategies to integrate LLMs into PBL, enhancing the learning experience of middle school students. 

\subsection{Design Considerations from an Exploratory Study}

In our exploratory study, aimed at identifying challenges in LLM-based PBL design, we focused on integrating LLMs into a PBL environment to enhance the educational experience of middle school students. This study involved 12 students, aged 14-16 (OS1-OS12, M = 15, SD = 0.58), guided by college-level mentors (OM1-OM3, M = 23.3, SD = 1.89; demographic details are provided in Appendix C.1). Within the seven-day PBL program, 12 students were divided into three groups (OG1-OG3), each tasked with developing innovative solutions for creating a low-carbon campus environment. Four researchers conducted detailed observations and tracked students' performance throughout the program. Additionally, the mentors assisted in observing and documenting the in-class behaviors and interactions. At the end of the project, we interviewed students and mentors. All interviews were recorded. The recordings were transcribed on Feishu\footnote{\url{https://www.feishu.cn/}}, a web tool used to automate speech-to-text transcription. Four researchers first read transcripts separately, then debriefed and discussed findings to reach an agreement in team meetings. Through data analysis, we concluded five challenges and proposed five design considerations for the LLM-based PBL design. The study was conducted with full ethical approval and informed consent. Additionally, the study was approved by our university's Institutional Review Board (IRB).

\subsubsection{Lack of Clear Steps to Integrate LLMs}
Through observation, it is found that in the implementation of PBL, each group lacked clearly defined stages for the steps and methods used to develop creative solutions. For example, OM2 stated, \textit{"During the program, we often don't know what we should do next. We are not very familiar with the process of creative PBL, so sometimes we forget to use LLMs."} Besides, OM1 stated, \textit{"This lack of structure makes it difficult for mentors to provide clear and timely guidance on how to use LLMs effectively."} Therefore, to improve the effect of PBL, it is recommended to use a concise design thinking method to structure the stages of PBL. This method helps to integrate LLMs into PBL promptly and effectively. Based on this, we developed our design consideration 1:  

\begin{quote}\textbf{Design Consideration 1: Clarifying the Process of PBL and Timely Introduction of LLMs for Creative Problem-Solving.}\end{quote}

\subsubsection{Unclear Purpose for LLMs Usage}
During the project, it was observed that due to the complicated problems solved by creative PBL and its complex processes, both students and mentors found it challenging to grasp the purpose at each stage, making it difficult to utilize LLMs effectively. OM2 stated, \textit{"Students often don't know how to use LLMs to support creativity at different program stages. Sometimes, they use them blindly."} This method encourages viewing situations and challenges from multiple perspectives, promotes innovative thinking, and transcends traditional thinking processes. Engaging some thinking tools may guide students to apply LLMs effectively at each program stage. This suggestion is based on the observation that students often face difficulties when using LLMs for specific tasks. Therefore, here comes our design consideration 2: 

\begin{quote}\textbf{Design Consideration 2: Introduce Thinking Tools to Enhance the Usage Efficiency of LLMs at Each Stage.} \end{quote}

\subsubsection{Fail to formulate prompts to LLMs} 
The observation reveals that many students struggle to write clear prompts when using LLMs, which leads to unsatisfactory responses. OS6 and OS9 said, \textit{"We've never used LLM tools before and are accustomed to using simple keywords for search. We're unfamiliar with formulating complete sentences for inquiries, making it challenging to receive ideal answers."} This issue often results in a repetitive cycle of trial and error, prompting students to regularly seek assistance from their mentors. As OM1 shared, \textit{"Students frequently struggle with formulating accurate prompts and questions and often resort to my help. They communicate their intended question to me, and I assist them in posing it to LLMs."} To prevent students with limited LLM experience from wasting time familiarizing themselves with these tools, it's recommended to provide basic training on LLMs usage principles before starting the program. This training should involve supervised practice under mentor supervision. Based on the above, we enacted our design consideration 3:

\begin{quote}\textbf{Design Consideration 3: Instructing Students in Fundamental LLMs Operational Principles.}\end{quote}

\subsubsection{Overintervention and Insufficient Guidance from Mentors in LLMs Use.}
During our exploratory study, we found that the roles and responsibilities of mentors are not clearly defined. Some mentors overstep, interfering with students' use of LLMs, while others do not offer enough guidance. For instance, OM1 tends to intervene when students face challenges, often directly assisting students in asking LLMs questions, leading to students' over-reliance. Conversely, OM3 offers minimal guidance, leaving students unsure about when they can seek help. Given the crucial role of mentors in enhancing the effective use of LLMs in PBL, a comprehensive mentor training plan is needed. The plan will clarify their roles, responsibilities, and the appropriate use of LLM tools in the educational process. Therefore, we developed our design consideration 4:

\begin{quote}\textbf{Design Consideration 4: Mentor Training and Support in LLMs Integration.} \end{quote}

\subsubsection{Challenges in Balancing LLMs Use and Group Collaboration} 

In our exploratory study, we did not limit how each group used LLM tools, allowing them to use LLMs freely. However, we observed that in some groups, several students were overly addicted to using LLMs, resulting in neglect of group discussions and collaboration. OM1 said, \textit{"One student in our group likes to use this tool. He just kept tinkering with the computer by himself, not participating in the group's collective discussion, which caused him not being able to keep up with the group's progress and not cooperating well."} To solve this problem, OM1 and OM2 suggested that each group could be assigned only one computer equipped with LLM tools to promote collective discussion. In addition, we used established LLM tools in this study, such as Bard, Bing Chat, and ChatGPT. However, students need a lot of professional knowledge support and timely feedback, especially for specific preferences for Chinese inquiries. Therefore, choosing a suitable LLM tool and determining its usage are very important. Thus, we formulated our design consideration 5:

\begin{quote}\textbf{Design Consideration 5: Determining the Mode and Tools of LLM Use} \end{quote}

These findings underscore the need for a well-structured approach to integrating LLMs into PBL. Clear guidelines,  mentor training, appropriate tool selection, and strategic use are paramount. These considerations support the subsequent design of an LLM-empowered PBL program, ensuring an optimized learning experience that leverages the capabilities of LLMs in educational settings.

\section{Experimental Study}
Based on the design considerations from the above comprehensive exploratory study, this section presents the design of a creative PBL program empowered by LLMs. We conducted an instructional experiment with mixed-method data collection and analysis.

\subsection{Program Design}
To conduct the experimental study that applies LLMs to creative PBL, we collaborated with three experienced teachers to structure the PBL program based on the above design considerations. The specific program design is as follows (details are provided in Appendix B):

\subsubsection{PBL Process Design: Streamlining PBL for Creative Problem-Solving (Consideration 1)}

In line with our first design consideration, we use a concise design thinking method, like the Double Diamond model~\cite{kochanowska2022double,tschimmel2012design}, to structure the stages of PBL. This method helps to integrate LLMs into PBL promptly and effectively. we structured the program into four distinct stages (see Fig. 1 for details). This approach was guided by the principle of integrating LLMs into the PBL process, ensuring their use was effective and conducive to enhancing students' creativity. 

\begin{figure}[h]
  \includegraphics[width=\textwidth]{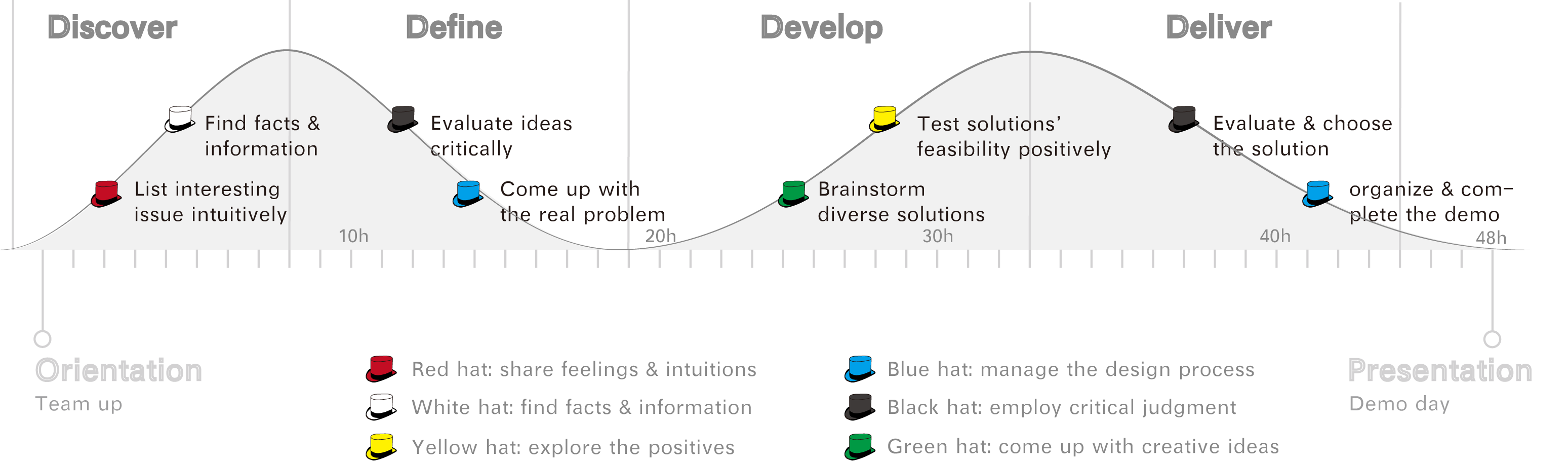}
  \caption{Six thinking hats in the PBL design process for LLMs usage }
  \label{fig:6hats}
\end{figure}

\textbf{Discover Stage (10h)}: Discover more real-life problems.

In the \dis, after equipping students with adequate knowledge about carbon emissions through lectures, students used a user journey map detailing the carbon footprint of an individual's daily activities, identifying carbon emission issues in their daily lives. They were encouraged to discover as many real-life problems as possible without considering the quality of the problems. During the process, students could utilize LLMs to seek information related to the problems they raised.

\textbf{Define Stage (6h)}: Refine and summarize the final problem.

In the \defi, students discussed, evaluated, filtered, and refined the previously identified problems. They were asked to consider the authenticity, significance, and feasibility of each problem, ultimately extracting a core problem for their group to focus on throughout the program. In this stage, students could employ LLMs like ChatGPT and Tiangong to acquire knowledge related to the questions or to validate the authenticity and validity of the problems.

\textbf{Develop Stage (10h)}: Boldly envisioning possible solutions.

In the \dev, students began by brainstorming solutions to their group's problems. They were encouraged to generate at least 15 ideas without limitations. They could use LLMs for information retrieval and idea generation during this process. After discarding uncreative or meaningless options, they evaluated the remaining for scientific validity, creativity, and potential use, using LLMs for feasibility checks and feedback. Finally, each group presented 3-5 solutions, receiving feedback from three experts in carbon emissions, design, and engineering.

\textbf{Deliver Stage (22h)}: Completing the final project.

In the \del, students refined solutions using LEGO Spike and Arduino knowledge. They chose an innovative and feasible solution to develop further, using LLMs for additional information and assessment of feasibility regarding the solution within the constraints of the learned technologies and available resources. After finalizing the solution, they began prototyping, during which they could use LLMs to overcome challenges. At the program's end, each team presented their problem and prototype, which five teachers evaluated.

\subsubsection{Thinking Tools Involved: Integrating Thinking Tools for Enhanced Application (Consideration 2)}

In our program, we involve De Bono’s ‘six thinking hats~\cite{kivunja2015using,goccmen2019effects}, representing different thinking directions respectively, which can significantly enhance the creative problem-solving and decision-making abilities of teams and individuals. This approach offers a new way to enhance the students’ creativity ~\cite{hu2021coordinating}. This method aligns with our second design consideration, emphasizing the importance of introducing thinking tools and clarifying LLMs functions in the PBL process.

\textbf{Discover Stage---White and Red Hats:} In the \dis, students utilized white hat thinking for gathering information and red hat thinking for emotional and intuitive insights. This combination encouraged a broad and optimistic exploration of issues related to the problem. LLMs served as a resource for information gathering, complementing the white hat's focus on data and facts.

\textbf{Define Stage---Blue and Black Hats:} In the \defi, blue hat thinking was employed for process control and organization, while black hat thinking was used for critical assessment of the problems. This stage required students to think analytically and practically, with LLMs providing support for validating the authenticity and feasibility of identified real-life problems.

\textbf{Develop Stage---Green and Yellow Hats:} In the \dev, green hat thinking fostered creative thinking for solution generation, and yellow hat thinking was used for evaluating the possibility and feasibility of these solutions. LLMs played a role in offering creative solution ideas and helping students assess their practicality.

\textbf{Deliver Stage---Blue and Black Hats:} In the \del, the blue hat's focus on process management and the black hat's critical judgment was crucial. Students refined their solutions and prototypes, with LLMs aiding in deepening solution development and evaluating feasibility within technological and resource constraints.

This structured application of the "Six Thinking Hats", coupled with the strategic use of LLMs, was designed to enhance students' critical and creative thinking skills and optimize the educational impact of LLMs in the creative PBL.

\subsubsection{Students Training: Principles for Effective LLMs Use (Consideration 3)}

We integrated a training session before the program to assist students with limited experience in LLMs. This was to familiarize students with LLMs, including practice sessions under guidance and supervision, guidance on framing questions and prompts, and integration of the responded information into their PBL tasks. By providing students with these skills, we aimed to streamline their learning process and enhance the overall efficiency of the creative PBL program.

\subsubsection{Mentor Training: Enhancing LLMs Guidance in PBL (Consideration 4)}

We implemented a specialized training program for mentors to facilitate the effective use of LLMs in our PBL. This program equipped mentors with the necessary skills for guiding students in using LLM tools, emphasizing their roles and guidelines. The focus areas include LLMs pre-setting, promoting collaboration through LLM tools allocation, and balancing mentor guidance with student autonomy. The program ensured all group members could access the LLM tools. Our research team provided ongoing observation and real-time support during the program to address any emerging needs in the classroom.

\subsubsection{LLMs Usage Design: Optimizing LLMs Choices and Methods for Collaboration (Consideration 5)}

Initial observations showed students using LLM tools individually, reducing group collaboration. We addressed this by providing shared LLM access points, with each group given one laptop with LLM tools. This promoted collaboration, dialogue, and collective decision-making. It also ensured that the use of LLMs was a joint effort rather than an individual task, fostering a collaborative environment and promoting group dynamics. In our experiment, we chose ChatGPT for its responsiveness to diverse queries and Tiangong\footnote{\url{https://www.tiangonglca.org/ai}}, a Tsinghua University-developed LLM tool for environmental education, which has changed the name to "Kaiwu" now. This ensured the tools were effective and relevant to sustainable development and carbon emissions.

\subsection{Procedures}

In our study, we conducted an instructional experiment framed as a week-long program featuring 48 hours, centered on "Low-Carbon Campus" using LLMs. Firstly, we initiated the project through a collaboration with a local high school, circulating information about the experiment and inviting students to participate. The registration form collected demographic details, previous experience with LLMs, and knowledge of low-carbon environments. Based on these submissions, we recruited students to participate in the experiment. Before the experiment, we informed the students and their parents about the study's objectives. Consent forms were obtained from both the students and their parents, ensuring ethical compliance and transparency. Additionally, the experiment was approved by our university's Institutional Review Board (IRB).

Secondly, we recruited seven mentors from university and graduate students with backgrounds in education, environmental science, and design technology. These mentors, sourced through social media, played a pivotal role in guiding the teaching process and assisting in our research by observing and recording the students’ interactions with LLMs. We divided students into seven groups (G1-G7), ensuring a mix of genders and levels of prior knowledge. Every group consists of four students and one mentor. The groups were tasked with identifying a low-carbon campus issue and developing a solution using LLMs, culminating in the creation of prototypes with tools like LEGO and Arduino. 

Thirdly, during the program, four researchers were actively involved in observing the students throughout the teaching sessions. We meticulously recorded each group's performance and interactions with LLMs. At the end of each day, we conducted discussions with the mentors to delve deeper into the students' experiences and feedback with LLMs. 

Finally, after the completion of the projects, we conducted semi-structured interviews with each student. These interviews aimed at capturing detailed experiences, were recorded and transcribed for thematic analysis, providing invaluable qualitative data for our study.

\subsection{Participants}
Our experimental study engaged a total of 31 students (S1-S31), aged between 15 and 16 years old (M = 15.06, SD = 0.36), randomly recruited from a local middle school. The demographic details of these students, including their gender, age, grade, prior experience with LLMs, and knowledge about low-carbon environments, are further elaborated in Appendix C.2. Additionally, our project was supported by seven mentors (M1-M7), consisting of graduate and undergraduate students. These mentors were selected for their diverse academic backgrounds, primarily in education, environmental science, and design technology(M = 23.1, SD = 2.03). They were critical in providing technical support and facilitating discussions among student groups. Detailed demographic information about the mentors is also available in Appendix C.2.

The student participants were divided into seven groups (G1-G7), each consisting of four to five students and one mentor. This grouping strategy was designed to ensure a balanced representation of skills, knowledge, and experience with LLMs and environmental concepts within each group. The primary goal of this structuring was to optimally integrate LLMs into different stages of the student's project work, fostering a collaborative and immersive learning experience.

\subsection{Data Collection and Analysis}
We used a mixed-method approach to explore the opportunities and challenges of integrating LLMs in creative PBL. This involved structured observations and comprehensive interviews with students and mentors.

\textbf{Interview.} 
To gain deeper insights into the participants' experiences and perceptions, we conducted extensive post-program interviews with all 31 students and seven mentors. These interviews were face-to-face and audio-recorded, lasting between 30 minutes to an hour, aligning with established qualitative research methodologies \cite{alshenqeeti2014interviewing, berg2001qualitativeresearch}. The semi-structured nature of these interviews was guided by key questions (detailed in Appendix A), ensuring comprehensive coverage of all research questions.

The process of analyzing these interviews involved an inductive thematic analysis method \cite{maguire2017doing, patton2014qualitative}. This analysis was performed by four researchers, who initially coded the transcripts at a sentence or paragraph level. The focus was on identifying patterns and themes that emerged naturally from the data that were relevant to our research questions. These emerging themes were then discussed in team meetings and categorized into three distinct groups, which are elaborately reported in our findings section. The comprehensive thematic analysis map is available in Appendix F. It is important to note that all interviews were conducted in Chinese, and the first author was responsible for translating these transcripts into English, ensuring the preservation of the nuanced meanings and contexts.

\textbf{Observation.} 

To comprehensively understand the dynamics of student interaction with LLMs, our data collection also included comprehensive observations. Both researchers and mentors actively observed and recorded the students’ interactions with the LLMs throughout the entire program. These observations were vital in capturing real-time usage, challenges, and the overall engagement of students with the LLMs.

Mentors use an app to track the performance of group members in PBL. They can post photos, videos, and notes of participants every day and leave comments on each other's posts (see Appendix D). At the end of each day, mentors and researchers discussed their findings and reflected on the student's performance throughout the day. All reflections and discussions were recorded for research use. In addition, they also reviewed the history of the LLMs to see if there were ways to encourage students to generate high-quality prompts and guide them to iterate (see Appendix E). 

In addition to the observations, our team held regular debriefing discussions with the mentors during the program. These discussions were recorded and transcribed into text, serving as supplementary qualitative data. These data provided invaluable insights into the students' LLM usage and highlighted critical considerations for future PBL initiatives. The records of these observations and discussions, along with the pan and paper used for tracking and recording, contributed significantly to our understanding of the ways in which LLMs can enhance the PBL experience.

\section{Findings}
\begin{figure}[h]
  \includegraphics[width=\textwidth]{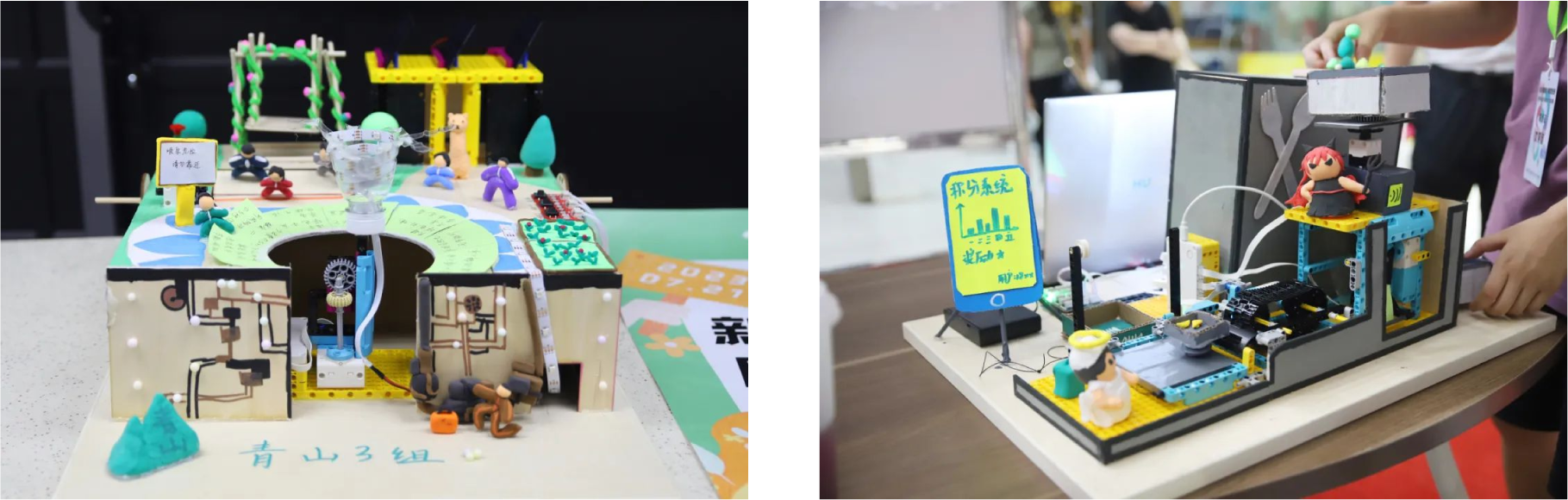}
  \caption{Final demo example of the G3 and G4 in the PBL program}
  \label{fig:outcome}
\end{figure}

Throughout the PBL program, students actively identified issues in their surroundings, defined the problems they aimed to address, proposed innovative solutions, and crafted creative prototypes using available materials, as illustrated in Fig.~\ref{fig:outcome}. This week-long program encompasses 48 hours of activities, culminating in students' successful completion of their PBL activities. Their final presentations were highly acclaimed, receiving unanimous approval from both peers and teachers. These results highlight LLMs' crucial role in modern educational frameworks and show how they can greatly improve student learning outcomes and experiences in PBL settings.

\subsection{LLMs Boosting Various Stages of Creative PBL}

\begin{figure}[h]
  \includegraphics[width=\textwidth]{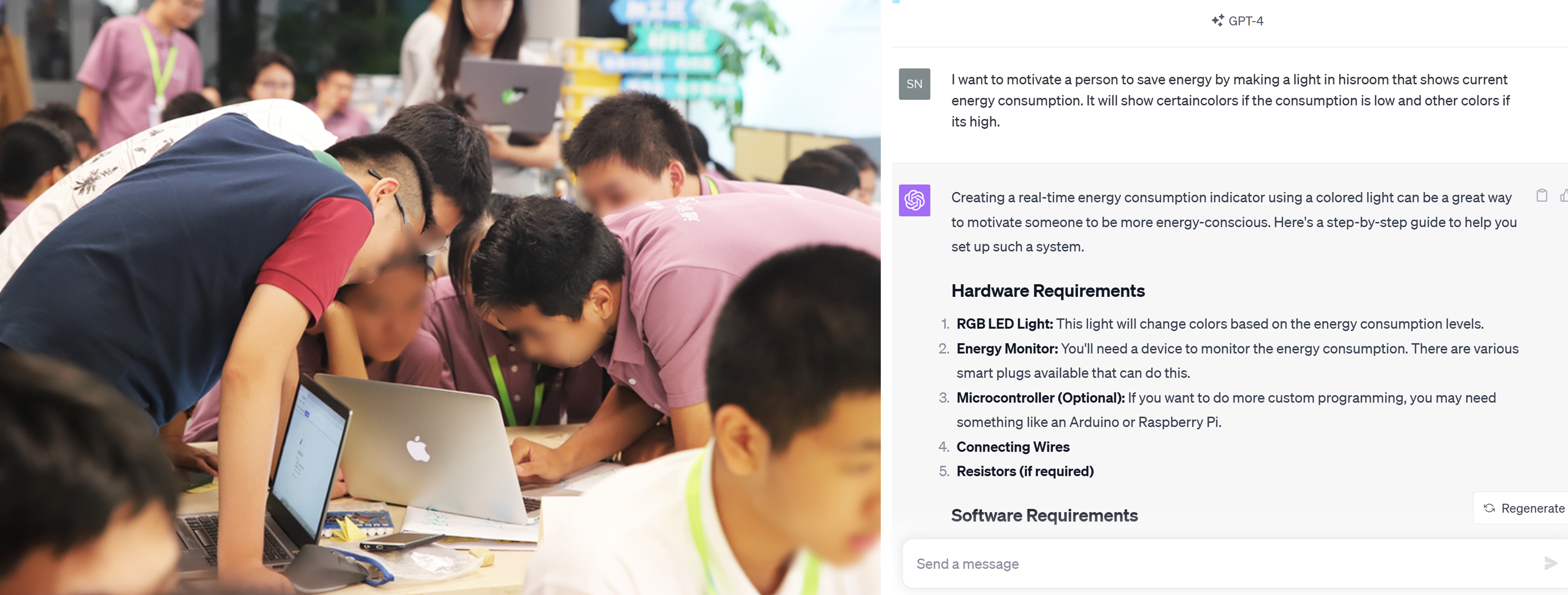}
  \caption{Students are collaborating with LLMs on their project}
  \label{fig:stucol}
\end{figure}

LLMs benefit from different stages of creative PBL and help students finish their tasks efficiently, based on the interviews with students and mentors. Throughout the project, LLMs could simultaneously respond to each student's questions and support them to be more creative (as shown in Fig.~\ref{fig:stucol}), which is hard to realize in a traditional classroom. Therefore, LLMs can bring unprecedented benefits to creative PBL in various stages as they can embrace every idea of each student.

\subsubsection{Discover information with a deep understanding}

In the \dis, students were provided with some structured and thoughtful answers through the LLMs, which helped them to be able to understand the new area systematically and logically. S30 noted, \textit{"The traditional search engine can only provide fragmented information. Nevertheless, ChatGPT and Tiangong can provide you with information with depth and structure"}. Compared to giving some web pages with high relevance and one or two sentences of answers that pop up automatically, the answers given by LLMs are generally more logical and organized. For example, when asking questions related to environmental protection with ChatGPT or Tiangong, every main subject from the answer has bullet points underneath.

\begin{figure}[h]
  \includegraphics[width=\textwidth]{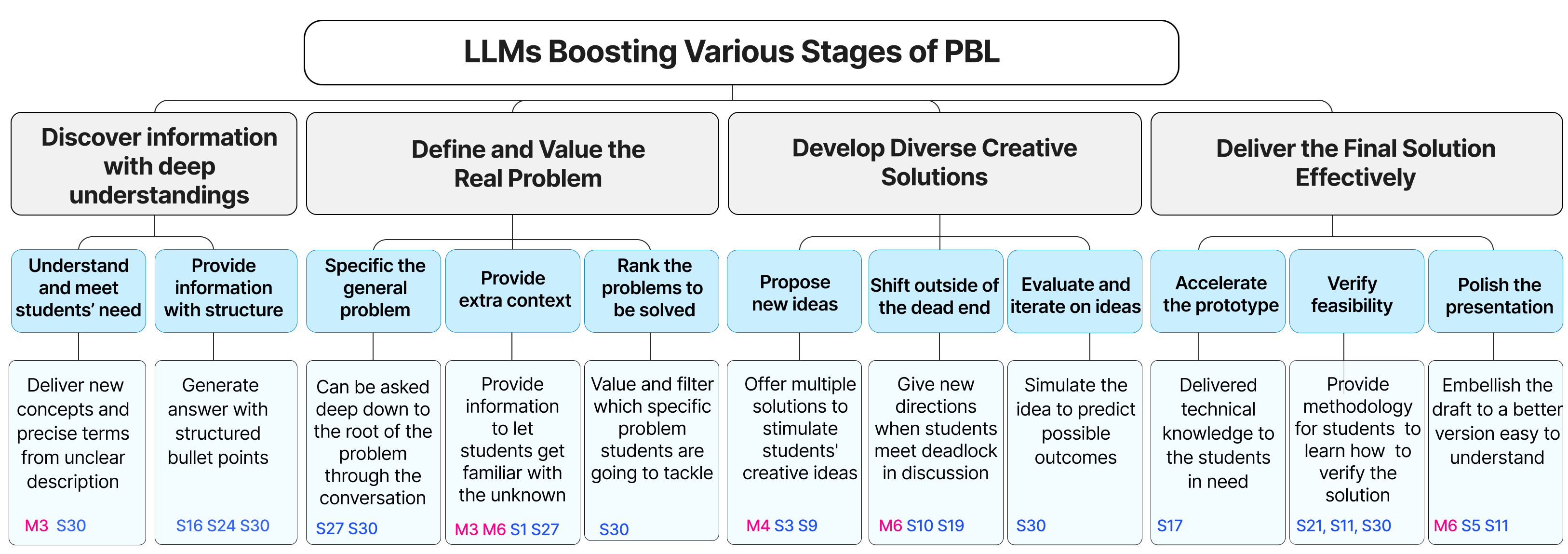}
  \caption{Thematic analysis map of LLMs boosting various stages of PBL}
  \label{fig:findings1}
\end{figure}

In addition, LLMs are extraordinarily understanding and meet students’ needs. Even if the student only has a rough idea, they have a powerful ability to help them discover the specific knowledge and concepts they expect. S24 said,
\textit{"Once I asked 'what physical quantity can describe how much light produces energy?' Tiangong gave me one accurate answer - photovoltaic conversion rate - and its specific explanations}". S16 and S30 had similar experiences and held the same viewpoint. In conclusion, LLMs can integrate key concepts from these vague descriptions and help students understand them quickly (as shown in Fig.~\ref{fig:findings1}).

\subsubsection{Define and Value the Real Problem}
In the \defi, students need to wrap up their ideas and find a real problem that needs to be solved. Since students lack relative experience and knowledge and are exposed to many new concepts at one time, they often come up with impractical problems that cannot be solved in a limited time. As a result, they rely heavily on the guidance of the teacher in traditional PBL. Thus, involving LLMs in PBL can provide personalized support and feedback that the teacher cannot achieve before.
Firstly, it helps students to define problems objectively with abundant, well-structured knowledge. S1 stated, \textit{"I used ChatGPT when defining a problem, such as checking some information to see if the problem is a real one and whether it has research value."} 
Secondly, students can gradually get to the root of the problem by interacting with LLMs. G6 focused on air condition issues in low-carbon campuses. Group member S27 stated,~\textit{"We guessed using air conditioning might cause lots of carbon emissions. So, we started asking about all kinds of emission sources, and it pointed out air conditioning after a few questions and proved our guess.}" 
In addition, students also used LLMs to evaluate and filter problems. For instance, S30 said, \textit{"We used ChatGPT for evaluation when deciding the research question between waste sorting system and energy-saving illuminating system. We finally pick the latter one based on its logic and creative answers"}. In summary, LLMs are powerful tools to help students identify the problem from vague illusions by providing structured background information, defining the problem, and evaluating the problem. 

\subsubsection{Develop Diverse Creative Solutions} 
In the \dev, LLMs can effectively stimulate students' creative inspiration during brainstorming sessions, helping them to propose new ideas. M4 reported his group's performance, \textit{"We chose to tackle the waste issue in the canteen. I clearly remember that during a brainstorming session, the team mostly lingered on waste management solutions and made no groundbreaking progress despite exhausting their ideas. However, I guided them using ChatGPT and Tiangong, and then discovered other potential solutions such as 'food packaging' and 'charity fridges,' which instantly broadened their thinking and led to the generation of more intriguing ideas."} In this situation, LLMs can address the problem of limiting the quality and quantity of design output caused by limited individual experience. Furthermore, due to its potential to foster creative thinking, LLMs can aid students in evaluating existing solutions and breaking away from deadlock. For example, M6 mentioned, 
\begin{quote}
“When students in our group are stuck in a heated phase due to challenging each other's solutions, I suggest they consider what LLMs might have to say at this moment. This strategy effectively breaks the deadlock and paves the way for a new round of discussions." 
\end{quote}
After generating sufficient creative solutions, students can use LLM to simulate and understand how to achieve them. As S30 said, \textit{"We used LLMs to conduct a pre-production or pre-exploration of our ideas, and then choose a final solution based on its plausibility and creativity.}  In conclusion, LLMs can serve as a valuable tool in this stage, enhancing solution generation and evaluation.

\subsubsection{Deliver Final Solution Effectively}
In the \del, LLMs can enhance the full process by offering technical information support, logical operations capabilities, and text polishing. 
On one hand, LLMs provide a fundamentally clear approach for students to refer to and further validate. For instance, S11 noted, \textit{"When we were trying to understand the power output per unit area of solar panels, the data provided by ChatGPT wasn't accurate. However, it gives us a reliable way to calculate, allowing us to revise and recalculate by ourselves."}  S22 and S33 similarly noted that it can aid them in more accurately assessing the feasibility of the solution. 
On the other hand, LLMs can assist in lowering cognitive difficulty by providing students with a clear reference approach when students face constraints due to the learned technologies and available resources. S17 mentioned, \textit{"We looked up how to use pressure sensors because the tutorial didn't cover it. We found that using LLMs to understand pressure sensors and other interfaces is extremely convenient."} The simultaneous feedback from LLMs makes the prototyping phase more productive and creative based on M7, 
\begin{quote}
    "LLMs would respond to you no matter what you say. In other words, the LLMs can provide immediate feedback, especially with positive and constructive ideas. The students can realize that their ideas are being taken seriously, which is the first step to encouraging creativity."
\end{quote}  Finally, during the presentation phase, LLMs can be used to embellish the draft. For instance, M6 said, \textit{"Our group used LLMs to polish after they drafting the presentation slide."}

In summary, the empowering potential of LLMs shows adequately in the whole PBL process. During the Discover and Define stage, students utilized them as a fast-track method to gain insights into a particular domain and validate the authenticity and relevance of their identified issues. Moving on to the Develop stage, students turned to LLMs for a more enriched brainstorming session. Lastly, during the Deliver stage, students often sought the assistance of LLMs to streamline their thoughts and construct a coherent narrative structure. In essence, the integration of LLMs in the creative PBL framework significantly enriched the learning journey, offering dynamic support at every stage and empowering students to complete their projects with enhanced confidence, precision, and creativity.

\subsection{Ambivalent Perspectives on LLMs in Creativity}
As we delve into the impact of LLMs like ChatGPT and Tiangong on creativity within PBL, students exhibit mixed attitudes. This section explores these varied perspectives, underscoring the complex interplay between technology and students' creativity.

\subsubsection{Better or Worse: Students' Views on the Impact of LLMs on Creativity}
In the process of using LLMs for creative PBL, students have varying experiences and opinions regarding the role and effects of LLMs in the divergent phase. Students with a positive attitude believe that LLMs catalyze their divergent thinking process, enabling them to surpass their knowledge boundaries and think outside the box. It empowers them to explore more creative problem statements and solutions. For instance, S4 noted, \textit{"During the brainstorming, LLMs provided us with some new terms and concepts that I had never met before, which is also very convenient for me to continue to ask and get deeper understandings."} S20 similarly mentioned that due to her limited knowledge base, she often struggled to come up with diverse ideas for a given problem. However, when she turned to ChatGPT, she expressed,  \textit{"It gave me a new perspective. It can generate a lot of feasible solutions, many of which I couldn't think of on my own. It can inspire me a lot."} Inspired by ChatGPT's responses, her team incorporated a screen displaying energy consumption data into their final project to raise awareness among students about turning off air conditioning in a low-carbon campus context. 

Similarly, S17 noted, "\textit{ChatGPT sometimes comes up with something that none of us had thought of, and it can be quite refined}". Initially, their team was only considering creating a leftover food weighing system as an indirect solution to the problem of food wastage among students. However, after interacting with ChatGPT, S17 mentioned, "\textit{One of the things ChatGPT suggested was refrigeration for food preservation, which we hadn't thought of. This opened up a new avenue for us, the leftovers could be reused and taken away}". ChatGPT's response inspired their team to explore the idea of a food packaging service, which is rarely provided in Chinese middle school cafeterias. This shifted their focus from a weighing and feedback system to a more direct approach to reducing food waste.

\begin{figure}[h]
  \includegraphics[width=\textwidth]{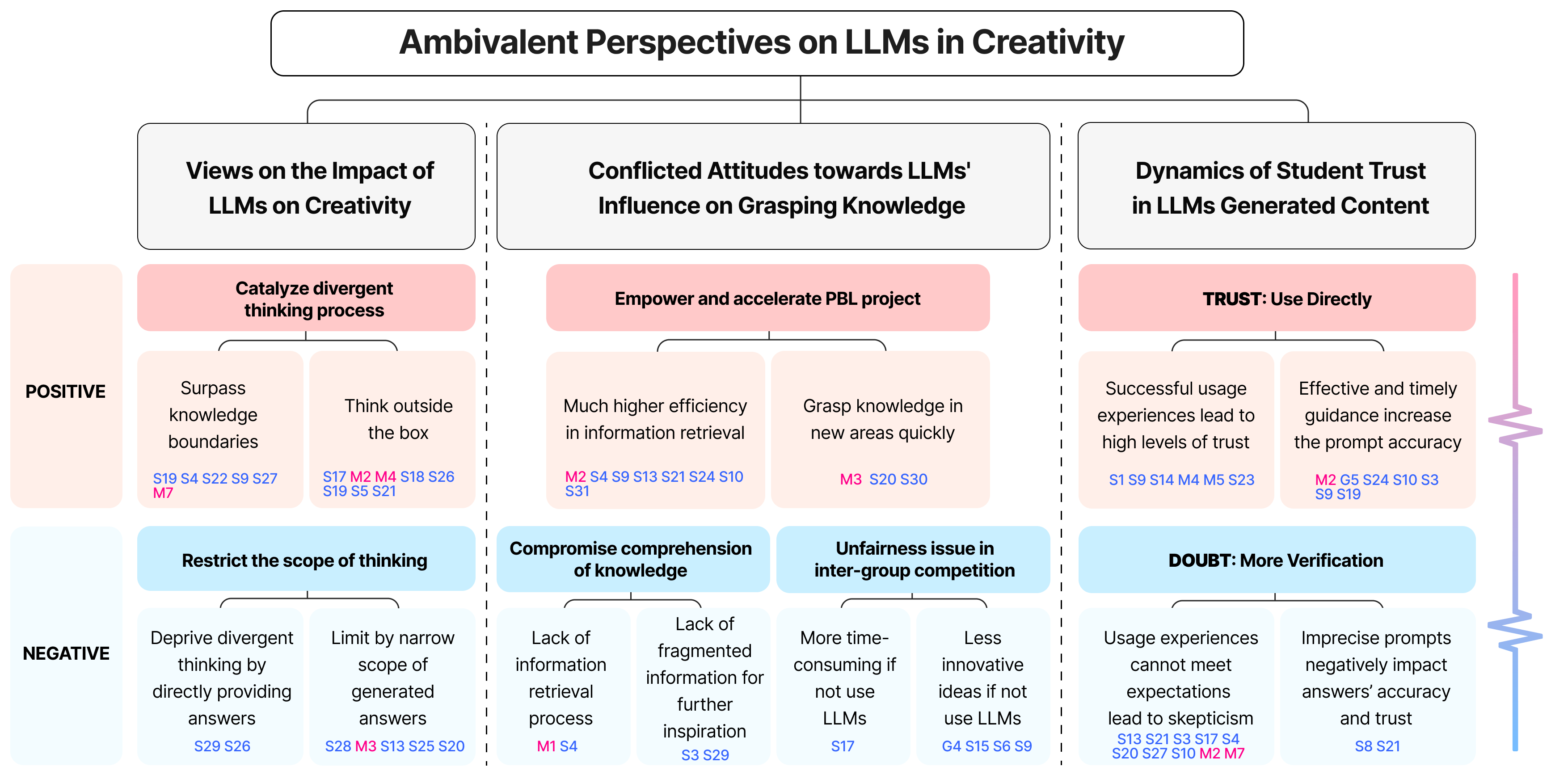}
  \caption{Thematic analysis map of ambivalent perspectives on LLMs in creativity}
  \label{fig:findings2}
\end{figure}

However, some students believe that ChatGPT did not contribute to their divergent thinking process and, in some cases, even restricted the scope of their thinking. S29 thought LLMs deprive his opportunities for divergent thinking by directly providing answers, \begin{quote}
    "I don't want to use LLMs. I prefer a blank canvas and don't want it to provide me with any templates that might limit my ideas. If I use LLMs to search first, its ideas could heavily influence my thinking, but I don't want to be influenced in that way."
\end{quote} S28 also expressed her feeling of being limited, "I think the scope of answers given by LLMs is a little narrow, and sometimes it limits my thinking." Therefore, students have mixed expectations and concerns about using LLMs for a creative generation. (as shown in Fig.~\ref{fig:findings2})

\subsubsection{Comprehension or Competitiveness: Students' Conflicted Attitudes towards LLMs' Influence on Grasping Knowledge}

Due to the iteration and promotion of LLMs, both students and educators are grappling with whether to use this technology in the learning process. Middle school students' information literacy and retrieval abilities are usually still in the developmental stage, with many leaning toward search engines as their primary means of information acquisition. The introduction of LLMs has precipitated shared apprehensions among teachers and students alike.

On the one hand, some concerns deviating from traditional information retrieval and acquisition methods may compromise the in-depth comprehension of knowledge. M1 concerns, \begin{quote}"Students access information quickly through using LLMs. Although it seems to save time, it deprives students of the opportunity to gradually explore and understand knowledge. Therefore, access to information is not the same as the real grasp of knowledge, and the lack of this process means the loss of understanding of the causes and consequences of knowledge."\end{quote}
S3 also states a similar perspective: compared with using LLMs to generate well-organized answers, she prefers to use traditional search engines because "\textit{The search results from traditional search engines might be fragmented, but they could occasionally inspire me with more thoughts and further exploration}". 
 
Conversely, not using such technology could cause one to lag behind the rapid pace of contemporary information retrieval, consequently affecting competitiveness. As the findings stated in 6.2, students can use LLMs effectively to empower every phase of a project due to LLM's high efficiency in information retrieval. They can quickly grasp knowledge in new areas with the help of LLMs, thereby accelerating group progress. However, this also has drawn the attention of some students who are concerned about the unfair issue of LLM usage in inter-group competition. S17 expressed her worry about the difference of time-consuming between using LLMs and purely brainstorming on their own, "\textit{because if other groups have LLMs to assist them generating creative ideas, our team might be at a disadvantage in the progress if we refuse to use it}". Concurrently, other G4 members also said that although they wanted to exhaust their ideas and not rely on LLM, this would be time-consuming and may lead to their outcome being inferior to others because they didn't use LLMs to provide them with more innovative ideas. It's evident that as students integrate LLM into their group tasks, there exists different complex thinking between fully exercising their innovative problem-solving capabilities and capitalizing on the competitive advantages the tool offers for the group.

\subsubsection{Believe or Doubt: Dynamics of Student Trust in LLMs Generated Content}

Students' levels of trust in LLM-generated content vary significantly. Some students believed in this content, while others may expressed a critical perspective. Besides, mentors also mentioned some students' trust in LLMs fluctuates with different experiences while using them. M2 stated,
\begin{quote}
    "When students found that LLM could not give them the answers they wanted, they would show a little doubt and felt that LLM was not as useful and helpful as they imagined. As I guided them to get the answers they wanted, they began to trust LLM more. And after several positive experiences, students may trust LLM blindly, and I need to remind them to check the answer critically."
\end{quote}

From the above, it can be seen that different levels of trust in LLMs have resulted from different usage experiences. In terms of the successful usage experiences, S1's trust in ChatGPT originated from his initial question about "\textit{When will the Chinese national football team win the FIFA World Cup?}", to which he received a very objective response that fully convinced him. As for the unsuccessful usage experiences, we found that they were not solely due to ambiguities or errors in LLMs-generated content, but also arose from imprecise prompts provided by the students. S8's unsuccessful query experience was caused by merely presenting "\textit{carbon emissions from waste paper disposal}" to ChatGPT without specifying any constraints, like the desired processing method or volume of waste paper. Even though ChatGPT later indicated the need for more detailed constraints in its response, S8 and her teammates didn't optimize their prompt. Conversely, when G5 was calculating the power consumption of the cross-slider in a device, with guidance from a mentor, the prompt was iterated several times. They achieved the desired results by continuously adjusting their phrasing and adding constraints, leading to a positive perception of the LLMs. 

These different levels of trust influence students' usage patterns, particularly when students are skeptical about the generated content. They tend to take the content more critically, pay more attention to the data generation process, and verify it when using LLMs. For example, S3 mentioned that she sometimes re-calculates the calculation results provided by ChatGPT. S21 opted for sources known for their reliability, stating, "\textit{I prefer Tiangong because it provides information with their reference.}". Regarding factual data, S17 sometimes uses common sense for initial assessment, while S4 stated that they usually search the questions again on Baidu~\footnote{\url{www.baidu.com}} (China’s leading search engines, like Google) to verify. Furthermore, M2 and M7 have observed that some students give up on using LLMs forever when they realize the inaccuracies in generated content.

In summary, students‘ attitude towards LLMs is complicated. In the context of innovation and divergent thinking, LLMs can sometimes be enlightening but can also act as a constraint. During group collaborations, students need to strike a balance between relying on technology and their own thinking. Moreover, the level of trust in LLMs varies widely depending on the students' interactions with them.

\subsection{Challenges in Integrating LLMs in Creative PBL}
Despite the success achieved in the application of LLMs within this PBL project, we encountered several challenges during the experimental process. In the following section, we will systematically review and analyze these challenges to provide a comprehensive understanding and to inform future research endeavors in this field.

\subsubsection{Hard to Formulate Effective Questions for LLMs}

During our experimental process, we discovered that students sometimes struggle with utilizing LLMs to ask questions. This difficulty arises from their unfamiliarity with how to prompt effectively and their habituated method of knowledge acquisition. This is because middle school students in the traditional Chinese education system often passively acquire information and are seldom encouraged to explore and pose questions actively. S20 mentioned,
\begin{quote}"I would find it troublesome to formulate questions. Especially if multiple factors are involved, I find it difficult to summarize all my confusions clearly into one question."\end{quote} S21 and S27 addressed the same problem.
M1 also affirmed this viewpoint, \textit{"In guiding students in information retrieval, I found they are more adept at using the fragmented keyword search method. Students need more effort to identify and frame effective questions that LLMs can understand better."} Meanwhile, we observed students were hesitant to try new tools due to the fear of asking the wrong questions and making mistakes.
In summary, while the use of LLMs as an additional tool in creative PBL has been generally effective, it requires that students be capable of formulating questions using LLMs. However, this still presents challenges for some students.

\subsubsection{Excessive Reliance on Mentors' Guidance}
Our observations underscored the critical importance of mentor guidance in the process of incorporating LLMs into creative PBL. Mentors play an essential role in guiding students on how to prompt effectively. M1 mentioned \begin{quote}"Students often find it challenging to deconstruct the intricate problems they encounter into specific, manageable queries. As a result, my role becomes crucial in guiding them to segment the problem and formulate questions in a logical order to achieve the desired outcome."\end{quote}
Beyond this, mentors are also required to encourage students to actively explore new areas using unfamiliar tools in questioning interactive ways. For example, M2 stated, 
\textit{"In the beginning, students were very frustrated and even wanted to give up on using it because they could not get the expected results from LLMs. However, with my guidance and persistent encouragement, we ultimately managed to obtain a relatively satisfactory answer using LLMs."}
These experiences shed light on the unique technical and emotional hurdles that middle school students face when using LLMs. This points to the necessity for a more substantial mentor presence and a greater emphasis on mentor guidance in the context of LLMs-enabled creative PBL.

\subsubsection{Uncertain LLMs' Role in Collaboration}
Our study found that assigning a fixed role to LLM in a collaborative group setting is challenging. During the exploratory study, there was no limitation on the number of LLMs available in one group, which resulted in some group members becoming engrossed in dialogues with LLM tools and neglecting group collaboration. M2 stated,\textit{"This can leave some group members in a relatively isolated state, especially when several group members all have this dissociative state, it can directly hinder the group's collaborative work."}

In the experimental study, we switched to a "one laptop with LLMs per group" model for information retrieval. This effectively improved the quality of group collaboration and the collaborative atmosphere but led to inequality among group members when using the computer due to varying levels of personal experience. For instance, we observed that in G3, all prompting and searching tasks were assigned to one group member who was experienced in using ChatGPT and capable of questioning in English accurately and expertly. Other group members' queries had to be mediated through this particular member, which somewhat diminished their freedom in information retrieval. 

As for the LLMs' role in collaboration, students and mentors hold different viewpoints. As M6 stated, \textit{"LLMs is an assistant in my group. Without it, the team might need a dedicated person to collect relevant information."} However, S14 said, \textit{"In my mind, LLMs are not only tools but also my partner. Sometimes, I would argue with them in some questions unconsciously."} Therefore, it is still a challenge to define a suitable role, as a group member's personal assistant, a virtual partner in the group, or any other possible roles in collaboration for LLMs to support creativity while balancing the group collaboration and members' equal access to LLMs.

\subsection{User Reflections of Design Considerations}

To further explore the design of Child-Centric AI Learning Environments, we conducted a detailed analysis of feedback from students and mentors on the five design considerations implemented in our program. 

\subsubsection{PBL Process Design Reflections}
During the program design process, we adopted the "Double Diamond model", dividing the PBL into four stages. Throughout this process, students were guided to utilize LLMs, and mentors provided positive affirmations. M1 mentioned, 
\begin{quote}"The use of this model made the entire learning phase clearer and more efficient. Students could also better recognize the problems that needed to be addressed at each stage while using GPT."\end{quote} 

Furthermore, mentors also mentioned that the integration of LLMs matched well with the design of these stages. M4 stated, \textit{"I think the Double Diamond Model, which involves four stages such as discovering problems, defining problems, developing solutions, and delivering solutions, integrates smoothly with AI. LLMs can offer significant help to students at each stage."} Meanwhile, some mentors pointed out that while the introduction of the "Double Diamond Model" has significant pedagogical value for PBL, it may not directly affect the use of LLMs. Therefore, in the design process, a deeper analysis is needed to distinguish which designs have a direct impact on the application of LLMs in PBL.

\subsubsection{Thinking Tools Design Reflections}
To better define how LLMs support students in their creative endeavors at each stage of the program, we introduced the "Six Thinking Hats" as a thinking tool. This tool stimulates students' creativity through the combination of different hats at each stage and guides the use of LLMs where each hat is involved. Mentors provided varied feedback on this design approach. M2 stated, \textit{"This design further refines the tasks for each stage, helping students to formulate more specific questions to input into the LLM, focusing more on solving the current problem."} Similarly, M6 mentioned, \begin{quote}"The different thinking hats allow students to pose questions from various perspectives, encouraging everyone to inquire from different standpoints. For example, the black hat can guide students to critically filter the answers provided by LLMs, while the green hat can encourage more positive exploration."\end{quote} 

 M1 and M5 also noted, \textit{"After students become familiar with these thinking hats, they would involuntarily think from different aspects when pondering questions and wouldn't verbalize their thoughts while using LLMs. Therefore, we are not entirely sure about the impact of the thinking hats on the use of LLMs."} Hence, the design of the Six Thinking Hats might directly stimulate creative thinking but did not realize that ChatGPT or Tiangong could be used as a tool for changing perspectives through natural language dialogue. The approach to questioning remained traditional and did not involve dialogue, lacking advanced Prompt use based on this tool.

\subsubsection{Mentor’s Training Reflections}

Regarding Mentor's Training, M1, M2, M4, and M6 reported significant benefits, stating, \begin{quote}"This training familiarized me with the basic methods and techniques of using LLMs, allowing me to anticipate the usage process and provide better guidance to students. Moreover, the training emphasized the need to critically approach the answers generated by LLMs, which made me more critical in using them."\end{quote} However, M3 suggested, \textit{"Mentor training should not just lecture but should include simulations based on real project cases for problem summarization, which would clarify the issues that can be raised."} Furthermore, M5 noted that the training on how mentors can scaffold students could be more explicit, specifically on how to guide, and to what extent. More professional training in educational psychology might be needed to enable mentors to be effective guides.

\subsubsection{ Students’ Training Reflections}

Mentors provided varied feedback on the brief training provided to students. Some mentors observed that the training encouraged students to dare to try new tools, shifting their perspective on LLMs from initial resistance to openness, and they also learned basic usage techniques. M1 and M3 noted that some students were still unable to shift their mindset from using traditional search tools to generating prompts after the training, requiring detailed guidance from mentors to engage in multi-turn dialogues with LLMs. \textit{"Sometimes, they would present their questions to me, and then I would help transform these into queries understandable by LLMs, input them, and then explain the retrieved information back to the students."} Therefore, more detailed demonstrations and practice are needed in student training to help students transition to a questioning mindset.

\subsubsection{LLMs Usage and Group Collaboration Reflections}
In terms of LLM usage, we provided support with one device per group, but the performance varied across groups. G1 and G2 were completely autonomous in their discussions, with mentors sensing the students' needs, proactively using LLMs to provide answer references, or guiding the group to formulate questions for the mentor to ask. In G4 and G6, one or two students primarily used the tool, with limited interaction among group members. G3 had one person operating the tool while others gathered around to observe the LLM's feedback; G5 had members taking turns using the device. These different modes of usage resulted in varying outcomes of collaborative creation(see Fig. 6 for details). M4 expressed that each student needs to have first-hand experience using the tool, as this might be more beneficial for them to understand the capabilities of LLMs and become more adept at using them. Especially for middle school students, mentors noted that both technical skills and critical thinking need to be guided and look forward to clearer definitions of team roles in the future. 

\begin{figure}[h]
  \includegraphics[width=\textwidth]{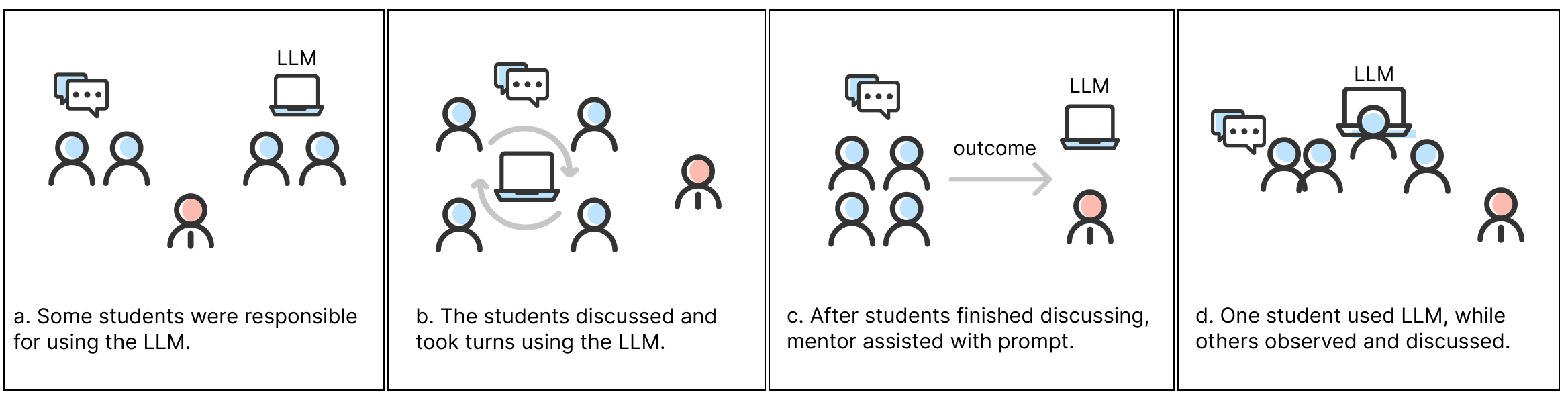}
  \caption{Four types of LLMs usage in the program }
  \label{fig:LLM Usage}
\end{figure}

\section{Discussion}

\subsection{LLMs' Positive Impact on Creative PBL}

The incorporation of LLMs into creative PBL has remarkably reshaped the educational landscape. Our research reveals that LLMs play a transformative role across various stages of PBL, from sparking initial curiosity to enhancing final presentations. The systematic information provision empowers students to acquire new domain knowledge essential for interdisciplinary creative projects rapidly. Furthermore, what's truly inspiring is the positive engagement fostered by LLMs, which quickly builds trust among students, enhancing their creative thinking and ability to think outside the box. This is particularly vital for completing creative PBL projects efficiently, especially under time constraints. 

A standout feature of LLMs is the ability to provide instant, personalized feedback that fills critical gaps in traditional educational settings, nurtures creativity, and improves information retrieval effectiveness. This aligns with some research that emphasizes the crucial role of feedback in the creative process, enhancing individual and group performance~\cite{wang2015enhancing,jung2010enhancing}. Wang, X. et al. underscore the value of informational feedback in computer-mediated idea generation, helping individuals to assess and improve their creative outputs objectively~\cite{wang2015enhancing}. Similarly, Jung et al. illustrated that in group settings, clear performance targets coupled with ongoing, objective feedback not only elevate motivation but also significantly improve overall performance~\cite{jung2010enhancing}. These studies indicate that timely and constructive feedback is essential for enhancing creative ideation and performance, a role effectively fulfilled by LLMs in educational contexts.

In conclusion, our study demonstrates that LLMs are not just tools for information dissemination but are pivotal in shaping a more interactive, responsive, and creative educational environment. The integration of LLMs into creative PBL signifies a shift towards a more dynamic, student-centered learning paradigm where technology and education converge to create an enriched, more effective learning experience. This integration is an exciting step towards redefining traditional educational models and preparing students for a rapidly evolving digital world.

\subsection{Concerns for LLMs in Education}
The integration of LLMs like ChatGPT in education has sparked a range of concerns among educators. A notable apprehension is the potential for LLMs to stifle creative thinking. The provision of ready-made answers by these models might limit students' inclination to engage in independent exploration and critical thinking~\cite{urban2023can}. This raises questions about the impact of LLMs on students' abilities to develop original ideas and engage in in-depth problem-solving processes. Additionally, the ease of accessing information through LLMs poses the risk of fostering an over-reliance on these tools. Such dependency could diminish students' critical thinking abilities and their capacity to evaluate the quality and accuracy of information~\cite{dwivedi2023so,yu2023reflection}. The instant availability of answers might also prevent students from pursuing comprehensive research or exploring topics in depth.

However, it's important to navigate these concerns with a balanced perspective. Embracing transformative technologies in a measured way can enable people to gradually accept them more informedly. For instance, integrating LLMs into the curriculum should come with comprehensive guidance on their effective use. This includes teaching students how to formulate impactful questions, critically analyze LLM-generated responses, and differentiate between high-quality and unreliable information. In addition to these considerations, the potential negative consequences of LLM outputs—such as biases, ethical issues, and the dissemination of harmful content—need thorough exploration. For example, educators can incorporate discussions and activities that highlight these issues, helping students recognize and address biases in AI-generated content. Ethical considerations, like the responsible use of information and respect for intellectual property, should be integral to the curriculum.

Recognizing and addressing these challenges is essential for a balanced assessment of LLMs in PBL. Rather than isolating students from these advanced technologies, the educational approach should focus on equipping them with the skills and critical awareness to use these tools effectively and responsibly. By doing so, we can harness the benefits of LLMs while mitigating their potential drawbacks, thus enriching the educational experience in line with evolving technological landscapes.

\subsection{Design Considerations and Implications for Effective LLMs Integration in Educational Settings}
Our study identified five design considerations from the exploratory study and applied them to the program design in the experimental study. Based on the analysis of user reflections on the program design, we derived five design implications for integrating LLMs into PBL.

\subsubsection{Defining a Structured Framework for LLM Integration in PBL}
Integrating LLMs into PBL needs a coherent, structured approach that not only ensures the effective application of these technologies but also enhances the educational experience by fostering creativity and critical thinking. To achieve this, incorporating a specific instructional framework can make the PBL phases more transparent, thereby clarifying the purpose of using LLMs at each stage. This structured approach should include comprehensive guidelines detailing the incorporation of LLMs for various tasks, such as brainstorming, research, problem-solving, and project execution. Crucial to this integration is the provision of extensive training for educators and students alike, covering LLM capabilities and limitations, ethical considerations, and interaction best practices. Equally important is the continuous evaluation and adaptation of the instructional model to accommodate the fast pace of AI advancements, ensuring that the use of LLMs remains relevant and aligned with educational objectives.

\subsubsection{Incorporating Thinking Tools for LLM Use in Creative PBL}
Incorporating thinking tools into PBL offers a strategy for guiding students toward more creative and effective use of LLMs. By intertwining tools like the "Six Thinking Hats" with LLM interactions, educators can foster an environment where students are encouraged to explore diverse perspectives and apply critical thinking at each stage of their projects. The integration of these tools necessitates clear, practical guidelines and examples that demonstrate how to effectively pair thinking strategies with LLM capabilities, such as using specific hats to direct queries or enhance solution generation through LLMs. Training sessions that emphasize the combined use of thinking tools and LLMs, featuring real-world scenarios and hands-on exercises, are essential for deepening students' understanding and proficiency. This approach not only bolsters the creative application of LLMs in educational settings but also ensures a reflective and critical engagement with the technology. Through dynamic incorporation of thinking tools alongside LLMs can significantly enhance both the learning experience and outcomes, preparing students to navigate and leverage the possibilities of AI with confidence and creativity.

\subsubsection{Developing Comprehensive Guidelines for Students to Trust and Effective Use of LLMs}
It is crucial to develop and integrate comprehensive guidelines to foster trust and enhance the usage experience of LLMs such as ChatGPT and Tiangong. These guidelines should aid students in understanding LLMs' fundamental functions and navigating their questioning strategies, illustrated through practical examples. For instance, providing case studies where LLMs were effectively used can help students understand the application of these tools under specific contexts. Additionally, emphasizes ethical and critical engagement with LLM-generated content, encouraging students to critically analyze and question the reliability and biases of AI-generated information.

\subsubsection{Customizing the Responsibilities and Guidance of Mentors}
Mentors play a crucial role in facilitating LLM integration, necessitating the customization of mentor responsibilities and guide methods. In the initial project phases, mentors could focus on helping students craft effective questions for LLMs. As projects progress, their role could evolve to include guiding students in interpreting and applying LLM responses in their work. For example, mentors can conduct workshops demonstrating different ways to utilize LLM outputs in various project stages, thereby ensuring that all students, irrespective of their specific project roles, can effectively leverage LLM capabilities. In the future, with the development of LLM technology, the development of agents specifically designed for educational scenarios can reduce the demand for mentors and alleviate the teaching burden.

\subsubsection{Establishing Clear Protocols for LLMs in Collaborative Environments}
The integration of LLMs into collaborative projects requires the establishment of clear protocols to define their roles and usage within groups. For instance, setting rules for shared access to ChatGPT can prevent any single team member from dominating the tool, thereby promoting equitable participation. Role rotation or scheduled LLM usage times could be implemented to ensure that every student has the opportunity to interact with the tool. This structured approach can maximize the benefits of LLMs in group settings while addressing potential challenges, such as over-reliance on the tool or unequal participation among team members. By clearly delineating LLMs' roles in line with team objectives and project requirements, their integration can be optimized to support, rather than overshadow, collaborative efforts.

\subsection{Limitation and Future Work}
Our study highlights the potential of LLMs within the PBL framework but also underscores several limitations that guide future research directions. Firstly, our study's limited generalizability is a notable limitation. We collaborated with a top-tier school in a major urban area, where students typically have high academic and creative aptitudes. Future studies should expand their reach to include a more diverse group of middle school students from various regions and academic backgrounds. This expansion will enable a more thorough and representative assessment of LLMs' impact in educational settings. 

By understanding the opportunities and challenges of involving LLMs in the classroom, future research could customize LLMs usage to individual creative needs and evaluate the actual effects of the customized LLM tools by conducting robust comparative experiments. Besides, future researchers also could expand the study's scope to gain a deeper and more inclusive understanding of LLMs in educational contexts.

\section{Conclusions}
Our research explored the integration of LLMs into creative PBL settings highlighting the transformative potential of LLMs in education. By conducting an instructional experiment in an authentic education scenario, we have unveiled the multifaceted implications of LLMs and provided evidence of their potential to boost the PBL process. Our findings reveal the vital function of LLMs across various PBL stages and highlight the complexity of student attitudes toward LLMs in creative tasks. We identified key challenges in integrating LLMs into creative middle school PBL, such as technology training, mentors' workload, and usage protocols. Additionally, our research addresses academic concerns about the potential drawbacks of LLM integration and offers valuable insights into the widespread and large-scale adoption of LLMs in the field of education. By addressing the challenges and leveraging the benefits of LLM-based PBL, we offer a nuanced perspective on how LLMs can be responsibly integrated into child-centric educational practices, paving the way for innovative, effective, and engaging learning methods. Our findings contribute to child-centered AI by offering practical insights into integrating LLMs in educational settings. 

\clearpage


\section*{Appendix}

\appendix 
\section{Interview protocol}

\subsection{Student's interview protocol}

1. Did you use LLMs in this program? At which stages did you choose to use, and which LLMs were used separately? (Tiangong, ChatGPT)\\
2. When using LLMs, what questions do you usually ask? \\
3. What is your most impressive experience of using generative LLMs? \\
4. Have you ever felt at a loss in using it? \\
5. Do you enjoy using LLMs to assist you in PBL learning? Why? Would you want to use these tools to help you in the future? 
\subsection{Mentor's interview protocol}
1. Did your group use LLMs in this program? At which stages did you choose to use, and which LLMs were used separately? (Tiangong, ChatGPT)\\
2. When using LLMs, what questions do your group usually ask? \\
3. What is your most impressive experience of using generative LLMs in your group? \\
4. Have your group ever felt at a loss in using it? \\
5. Do you enjoy using LLMs to assist you in PBL learning? Why? Would you want to use these tools to help you in the future? 

\section{The schedule of PBL for middle school students}
\begin{table}[]
\caption{The schedule of PBL for middle school students}
\begin{tabular}{llllllll}
\hline
                                                          & Sunday                                                                                     & Monday                                                                                        & Tuesday                                                                                                     & Wednesday                                                                                     & Thursday                                                                                               & Friday                                                                                                  & Saturday                                                                                               \\ \hline
\begin{tabular}[c]{@{}l@{}}9:00\\ -\\ 10:30\end{tabular}  & \begin{tabular}[c]{@{}l@{}}Opening\\ ceremony\end{tabular}                                 & \begin{tabular}[c]{@{}l@{}}Lecture\\ about\\ the usage\\ of LLMs\end{tabular}                 & \begin{tabular}[c]{@{}l@{}}Problems\\ defining\\ (supporting\\ by LLMS)\end{tabular}                        & \begin{tabular}[c]{@{}l@{}}Develop\\ solutions\\ (brain-\\ storming\\ with LLMs)\end{tabular} & \multirow{2}{*}{\begin{tabular}[c]{@{}l@{}}Prototype\\ delivering\\ (boosting\\ by LLMs)\end{tabular}} & \multirow{2}{*}{\begin{tabular}[c]{@{}l@{}}Prototype\\ delivering\\ (boosting\\ by LLMs)\end{tabular}}  & \multirow{2}{*}{\begin{tabular}[c]{@{}l@{}}Prototype\\ delivering\\ (boosting\\ by LLMs)\end{tabular}} \\ \cline{1-5}
\begin{tabular}[c]{@{}l@{}}10:40\\ -\\ 12:10\end{tabular} & \begin{tabular}[c]{@{}l@{}}Lecture\\ about\\ low carbon\\ campus\end{tabular}              & \begin{tabular}[c]{@{}l@{}}Subject\\ discovering\\ (supporting\\ by LLMs)\end{tabular}        & \begin{tabular}[c]{@{}l@{}}College \\ field\\ Environ-\\ mental\\ research\end{tabular}                     & \begin{tabular}[c]{@{}l@{}}Prototype\\ delivering\\ (boosting\\ by LLMs)\end{tabular}         &                                                                                                        &                                                                                                         &                                                                                                        \\ \hline
\begin{tabular}[c]{@{}l@{}}13:30\\ -\\ 15:00\end{tabular} & \multirow{2}{*}{\begin{tabular}[c]{@{}l@{}}Campus\\ orientation\\ \&Grouping\end{tabular}} & \multirow{2}{*}{\begin{tabular}[c]{@{}l@{}}Lecture\\ about\\ research\\ methods\end{tabular}} & \multirow{2}{*}{\begin{tabular}[c]{@{}l@{}}Lecture \\ about\\ the usage\\ of creative\\ tools\end{tabular}} & \begin{tabular}[c]{@{}l@{}}Test\\ viability\\ (supporting\\ by LLMs)\end{tabular}             & \multirow{2}{*}{\begin{tabular}[c]{@{}l@{}}Critique\\ from\\ experts\end{tabular}}                     & \multirow{2}{*}{\begin{tabular}[c]{@{}l@{}}Prototype\\ delivering\\ (boosting \\ by LLMs)\end{tabular}} & \begin{tabular}[c]{@{}l@{}}Presentation\\ \& Closing\\ ceremony\end{tabular}                           \\ \cline{1-1} \cline{5-5} \cline{8-8} 
\begin{tabular}[c]{@{}l@{}}15:20\\ -\\ 17:00\end{tabular} &                                                                                            &                                                                                               &                                                                                                             & \begin{tabular}[c]{@{}l@{}}Feasibility\\ report\end{tabular}                                  &                                                                                                        &                                                                                                         & \begin{tabular}[c]{@{}l@{}}Presenting\\ awards\end{tabular}                                            \\ \hline
\end{tabular}
\end{table}

\section{Demographic information of participant students}

\subsection{Exploratory study}
\begin{table}[]
\caption{Demographic information of participants in exploratory study}
\begin{tabular}{cccccc}
\hline
Group                & Participant & Gender & Age & \begin{tabular}[c]{@{}c@{}}Previous experience\\ to\\ LLM\end{tabular} & \begin{tabular}[c]{@{}c@{}}Previous experience\\ to \\ environment knowledge\end{tabular} \\ \hline
\multirow{5}{*}{OG1} & OM1         & F      & 26  & Y                                                                      & N                                                                                         \\ \cline{2-6} 
                     & OS1         & M      & 15  & N                                                                      & Y                                                                                         \\ \cline{2-6} 
                     & OS3         & M      & 14  & N                                                                      & Y                                                                                         \\ \cline{2-6} 
                     & OS2         & M      & 15  & Y                                                                      & N                                                                                         \\ \cline{2-6} 
                     & OS4         & F      & 15  & N                                                                      & N                                                                                         \\ \hline
\multirow{5}{*}{OG2} & OM2         & M      & 22  & N                                                                      & Y                                                                                         \\ \cline{2-6} 
                     & OS5         & M      & 14  & N                                                                      & N                                                                                         \\ \cline{2-6} 
                     & OS6         & F      & 15  & N                                                                      & Y                                                                                         \\ \cline{2-6} 
                     & OS7         & F      & 15  & Y                                                                      & N                                                                                         \\ \cline{2-6} 
                     & OS8         & M      & 15  & N                                                                      & Y                                                                                         \\ \hline
\multirow{5}{*}{OG3} & OM3         & F      & 22  & Y                                                                      & Y                                                                                         \\ \cline{2-6} 
                     & OS9         & M      & 15  & Y                                                                      & Y                                                                                         \\ \cline{2-6} 
                     & OS10        & M      & 15  & Y                                                                      & N                                                                                         \\ \cline{2-6} 
                     & OS11        & M      & 16  & N                                                                      & N                                                                                         \\ \cline{2-6} 
                     & OS12        & F      & 16  & N                                                                      & Y                                                                                         \\ \hline
\end{tabular}
\end{table}
\clearpage

\subsection{Experimental Study}
\begin{table}[]
\begin{tabular}{cccccc}
\hline
Group               & Participant & Gender & Age & \begin{tabular}[c]{@{}c@{}}Previous experience\\ to\\ LLM\end{tabular} & \begin{tabular}[c]{@{}c@{}}Previous experience\\ to\\ environment knowledge\end{tabular} \\ \hline
\multirow{6}{*}{G1} & M1          & M      & 22  & Y                                                                      & N                                                                                        \\ \cline{2-6} 
                    & S1          & M      & 15  & Y                                                                      & N                                                                                        \\ \cline{2-6} 
                    & S2          & M      & 15  & N                                                                      & N                                                                                        \\ \cline{2-6} 
                    & S3          & F      & 15  & Y                                                                      & Y                                                                                        \\ \cline{2-6} 
                    & S4          & F      & 15  & Y                                                                      & Y                                                                                        \\ \cline{2-6} 
                    & S5          & F      & 15  & N                                                                      & Y                                                                                        \\ \hline
\multirow{5}{*}{G2} & M2          & F      & 22  & Y                                                                      & Y                                                                                        \\ \cline{2-6} 
                    & S6          & M      & 15  & N                                                                      & Y                                                                                        \\ \cline{2-6} 
                    & S7          & M      & 15  & N                                                                      & N                                                                                        \\ \cline{2-6} 
                    & S8          & M      & 15  & N                                                                      & Y                                                                                        \\ \cline{2-6} 
                    & S9          & F      & 15  & Y                                                                      & Y                                                                                        \\ \hline
\multirow{6}{*}{G3} & M3          & F      & 20  & Y                                                                      & Y                                                                                        \\ \cline{2-6} 
                    & S10         & M      & 15  & N                                                                      & Y                                                                                        \\ \cline{2-6} 
                    & S11         & F      & 15  & Y                                                                      & Y                                                                                        \\ \cline{2-6} 
                    & S12         & F      & 15  & N                                                                      & N                                                                                        \\ \cline{2-6} 
                    & S13         & M      & 15  & N                                                                      & N                                                                                        \\ \cline{2-6} 
                    & S14         & M      & 15  & Y                                                                      & Y                                                                                        \\ \hline
\multirow{6}{*}{G4} & M4          & M      & 26  & Y                                                                      & N                                                                                        \\ \cline{2-6} 
                    & S15         & M      & 15  & N                                                                      & Y                                                                                        \\ \cline{2-6} 
                    & S16         & M      & 15  & N                                                                      & N                                                                                        \\ \cline{2-6} 
                    & S17         & F      & 15  & N                                                                      & N                                                                                        \\ \cline{2-6} 
                    & S18         & M      & 17  & Y                                                                      & N                                                                                        \\ \cline{2-6} 
                    & S19         & F      & 15  & N                                                                      & N                                                                                        \\ \hline
\multirow{5}{*}{G5} & M5          & F      & 24  & Y                                                                      & Y                                                                                        \\ \cline{2-6} 
                    & S20         & M      & 15  & N                                                                      & N                                                                                        \\ \cline{2-6} 
                    & S21         & F      & 15  & Y                                                                      & N                                                                                        \\ \cline{2-6} 
                    & S22         & F      & 15  & Y                                                                      & N                                                                                        \\ \cline{2-6} 
                    & S23         & M      & 15  & N                                                                      & N                                                                                        \\ \hline
\multirow{5}{*}{G6} & M6          & F      & 25  & Y                                                                      & Y                                                                                        \\ \cline{2-6} 
                    & S24         & M      & 15  & Y                                                                      & N                                                                                        \\ \cline{2-6} 
                    & S25         & M      & 15  & N                                                                      & N                                                                                        \\ \cline{2-6} 
                    & S26         & F      & 15  & N                                                                      & N                                                                                        \\ \cline{2-6} 
                    & S27         & F      & 15  & N                                                                      & N                                                                                        \\ \hline
\multirow{5}{*}{G7} & M7          & F      & 23  & Y                                                                      & Y                                                                                        \\ \cline{2-6} 
                    & S28         & F      & 15  & Y                                                                      & N                                                                                        \\ \cline{2-6} 
                    & S29         & M      & 15  & Y                                                                      & N                                                                                        \\ \cline{2-6} 
                    & S30         & M      & 15  & Y                                                                      & N                                                                                        \\ \cline{2-6} 
                    & S31         & F      & 15  & N                                                                      & N                                                                                        \\ \hline
\end{tabular}
\caption{Demographic information of participants in experiment study}
\end{table}
\clearpage

\section{The Screenshot of Tracking Application}
\begin{figure}[h]  \includegraphics[width=\textwidth]{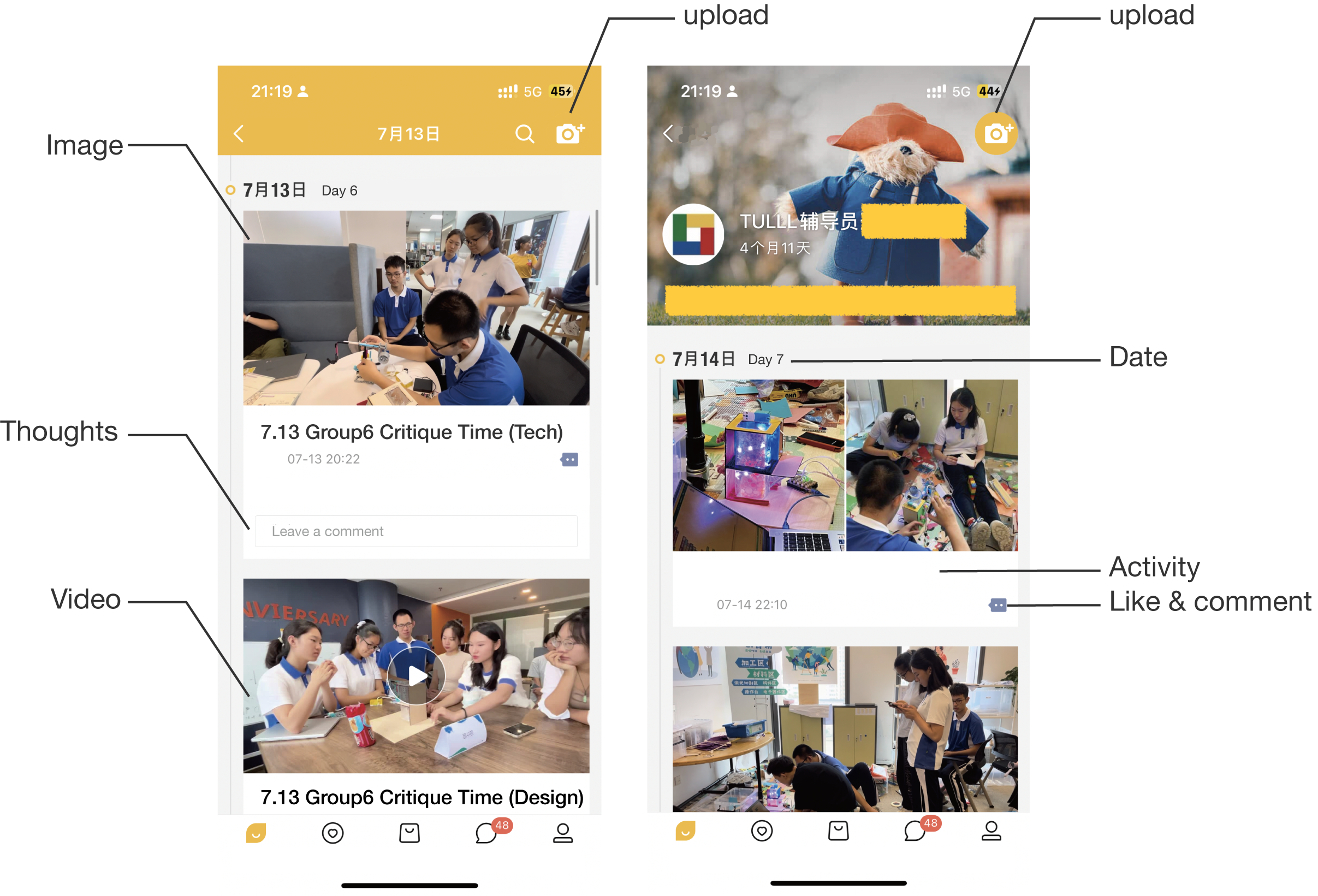}
  \caption{Interfaces of the application used for tracking project progress and personal performance}
  \label{fig:TrackingAPP}
\end{figure}
\clearpage

\section{The Historical Screenshots of LLM Tools }
  \begin{figure}[h]  
    \includegraphics[width=\textwidth]{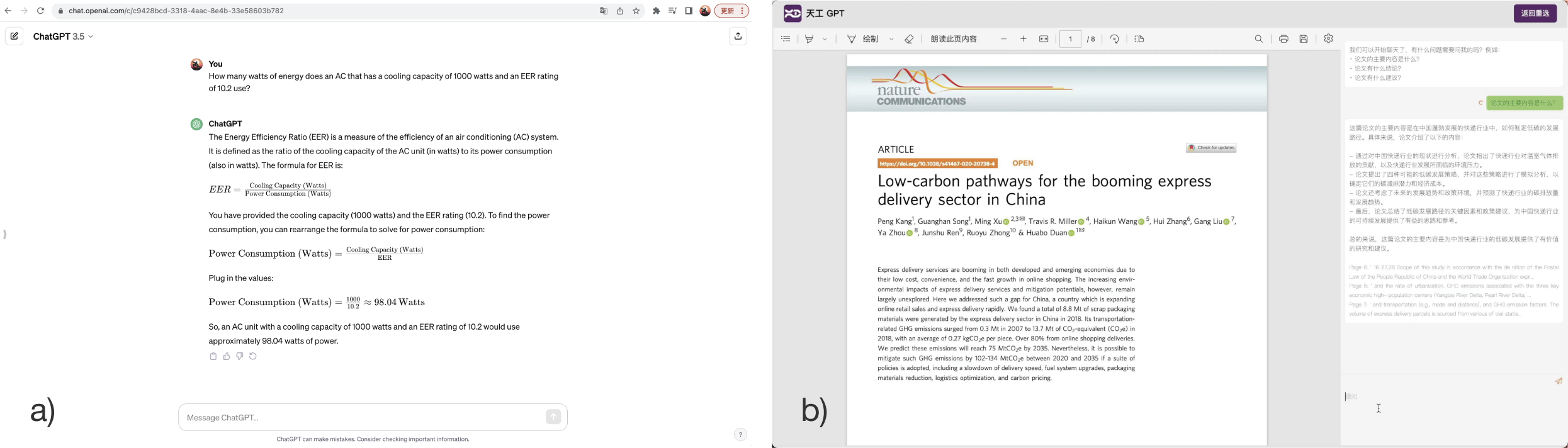}
    \caption{LLMs' screenshots: a) ChatGPT;  b) TianGong}
    \label{fig:LLMScreenshot}
  \end{figure}
  
\section{Thematic analysis map of findings}
  \begin{figure}[h]  
    \includegraphics[width=\textwidth]{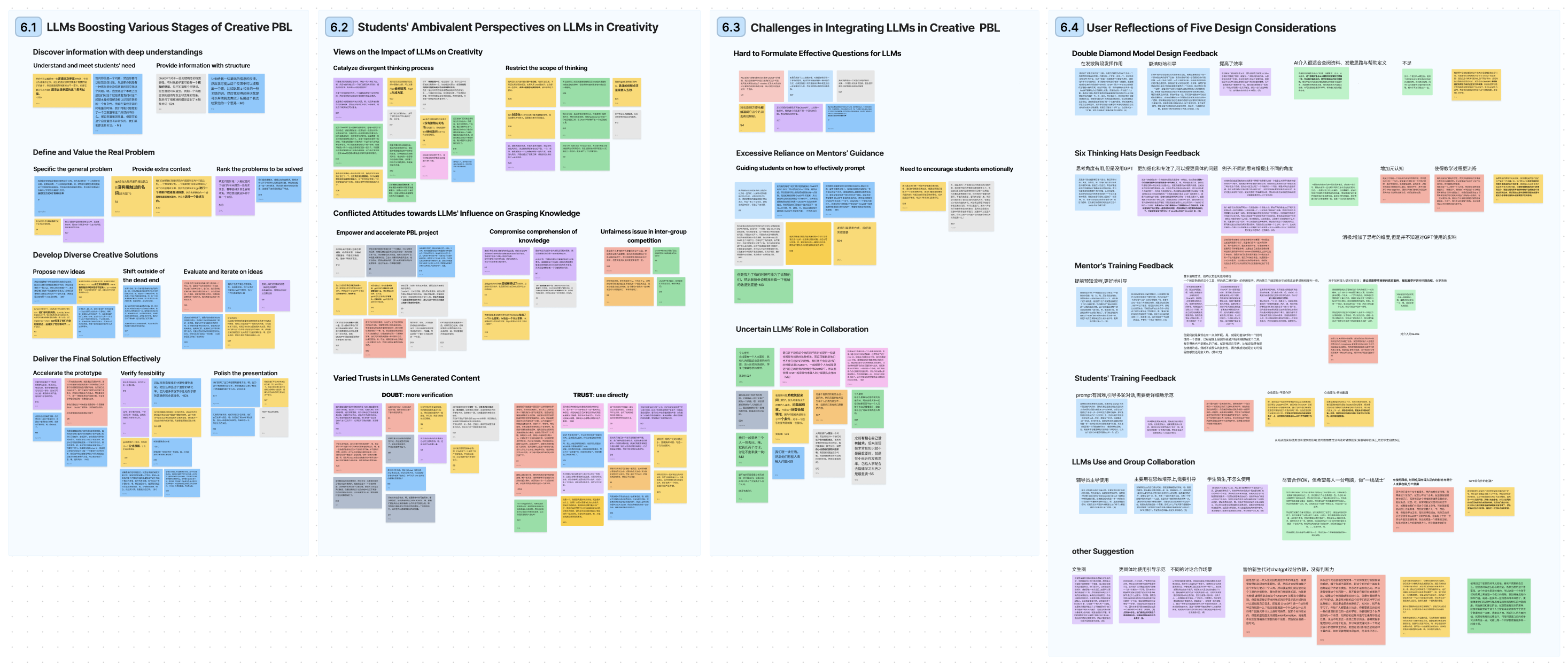}
    \caption{Theme analysis of findings.}
    \label{fig:text}
  \end{figure}

\clearpage

 \bibliographystyle{elsarticle-num} 
 \bibliography{main}
 
\end{document}